\documentclass[conference]{IEEEtran}
\IEEEoverridecommandlockouts

\usepackage{cite}
\usepackage{listings}
\usepackage{array}
\usepackage{balance}
\usepackage{multirow}
\usepackage{pbox}
\usepackage{url}
\usepackage{paralist}
\usepackage{xspace}
\usepackage{booktabs}
\usepackage{amsmath,amssymb,amsfonts}
\usepackage{algorithmic}
\usepackage{graphicx}
\usepackage{subfigure}
\usepackage{textcomp}
\usepackage{todonotes}
\usepackage{xcolor}
\usepackage{balance}
\usepackage{booktabs}

\newcommand{\pp}[1]{\vspace{3pt}\noindent\textbf{\emph{#1 --}}\xspace}

\begin{document}

\title{
WfBench: Automated Generation of\\Scientific Workflow Benchmarks
\thanks{This manuscript has been authored in part by UT-Battelle, LLC, under contract DE-AC05-00OR22725 with the US Department of Energy (DOE). The publisher, by accepting the article for publication, acknowledges that the U.S. Government retains a non-exclusive, paid up, irrevocable, world-wide license to publish or reproduce the published form of the manuscript, or allow others to do so, for U.S. Government purposes. The DOE will provide public access to these results in accordance with the DOE Public Access Plan (http://energy.gov/downloads/doe-public-access-plan).}
}

\author{
    \IEEEauthorblockN{
        Tainã Coleman\IEEEauthorrefmark{1}, 
        Henri Casanova\IEEEauthorrefmark{2}
        Ketan Maheshwari\IEEEauthorrefmark{3}, 
        Loïc Pottier\IEEEauthorrefmark{1}, 
        Sean R. Wilkinson\IEEEauthorrefmark{3}\\
        Justin Wozniak\IEEEauthorrefmark{4},
        Frédéric Suter\IEEEauthorrefmark{3},
        Mallikarjun Shankar\IEEEauthorrefmark{3},
        Rafael Ferreira da Silva\IEEEauthorrefmark{3}
    }
    \IEEEauthorblockA{
        \IEEEauthorrefmark{1}University of Southern California, Marina del Rey, CA, USA \ \ \ \ \
        \IEEEauthorrefmark{2}University of Hawaii, Honolulu, HI, USA \\
        \IEEEauthorrefmark{4}Argonne National Laboratory, Lemont, IL, USA \ \ \ \ \ \
        \IEEEauthorrefmark{3}Oak Ridge National Laboratory, Oak Ridge, TN, USA
    }
}

\maketitle

\begin{abstract}
The prevalence of scientific workflows with high computational demands 
calls for their execution on various distributed computing platforms, 
including large-scale leadership-class high-performance computing (HPC) 
clusters. To handle the deployment, monitoring, and optimization of 
workflow executions, many workflow systems have been developed
over the past decade. There is a need for workflow benchmarks that can 
be used to evaluate the performance of workflow systems on current and 
future software stacks and hardware platforms.

We present a generator of realistic workflow benchmark specifications that 
can be translated into benchmark code to be executed with 
current workflow systems. Our approach generates workflow tasks 
with arbitrary performance characteristics (CPU, memory, and I/O 
usage) and with realistic task dependency structures based on those 
seen in production workflows.  We present experimental 
results that show that our approach generates benchmarks that are 
representative of production workflows, and conduct a case study to 
demonstrate the use and usefulness of our generated benchmarks to
evaluate the performance of workflow systems under different configuration
scenarios.

\end{abstract}

\begin{IEEEkeywords}
scientific workflows, workflow benchmarks, distributed computing
\end{IEEEkeywords}

\section{Introduction}
\label{sec.intro}

Scientific workflows have supported some of the most significant discoveries 
of the past several decades~\cite{badia2017workflows} and are executed in
production daily to serve a wealth of scientific domains. Many workflows 
have high computational and I/O demands that warrant execution on large-scale 
parallel and distributed computing platforms. Because of the difficulties 
involved in deploying, monitoring, and optimizing workflow executions on 
these platforms, the past decade has seen a dramatic surge of workflow  systems~\cite{workflow-systems}.

Given the diversity of production workflows, the range of execution 
platforms, and the proliferation of workflow systems, it is crucial to 
quantify and compare the levels of performance that can be delivered to 
workflows by different platform configurations, workflow systems, and 
combinations thereof. As a result, the workflows community has recently 
recognized the need for workflow benchmarks~\cite{ferreiradasilva2021works}.
In this paper, we present a generator of realistic workflow benchmark
specifications that can be translated into benchmark code to
be executed with current workflow systems.

\subsection{Motivation}

Application benchmarks have long been developed for the purpose of
identifying performance bottlenecks and comparing HPC platforms.
Benchmarks have been developed that stress various aspects
of the platform (e.g., speed of integer and floating point
operations, memory, I/O, and network latency and throughput) and several
developed benchmark suites have become popular and are commonly
used~\cite{bailey1991parallel, dongarra2003linpack, hpcchallenge, 
ferenbaugh2015pennant, van2014parallel, slaughter2020task}.
A few of these benchmarks capture some, but not all, of the relevant 
features of production workflow applications: (i)~A workflow typically 
comprises tasks of many different ``types", i.e., that correspond to 
computations with different I/O, CPU, GPU, and memory 
consumption~\cite{ramakrishnan2008survey, ferreiradasilva-fgcs-2017}. 
(ii)~Even tasks of the same type, i.e., that are invocations of the same 
program or function, can
have different resource consumption based on the workflow configuration
(e.g., input parameters, input dataset). (iii)~In practice, several
workflow tasks are often executed concurrently on a single
compute node, causing performance interference and exacerbating
points~(i) and~(ii) above, which impacts workflow
execution time. (iv)~In production, workflows are executed using 
systems that orchestrate their execution and that can
be configured in various ways (e.g., task scheduling
decisions); Thus, it is crucial for workflow benchmarks to be
seamlessly executable using a wide range of these systems rather than being
implemented using one particular runtime system (e.g., as is the case for
classical HPC benchmarks implemented with MPI).

\begin{figure}[!t]
  \centering
  \includegraphics[width=\linewidth]{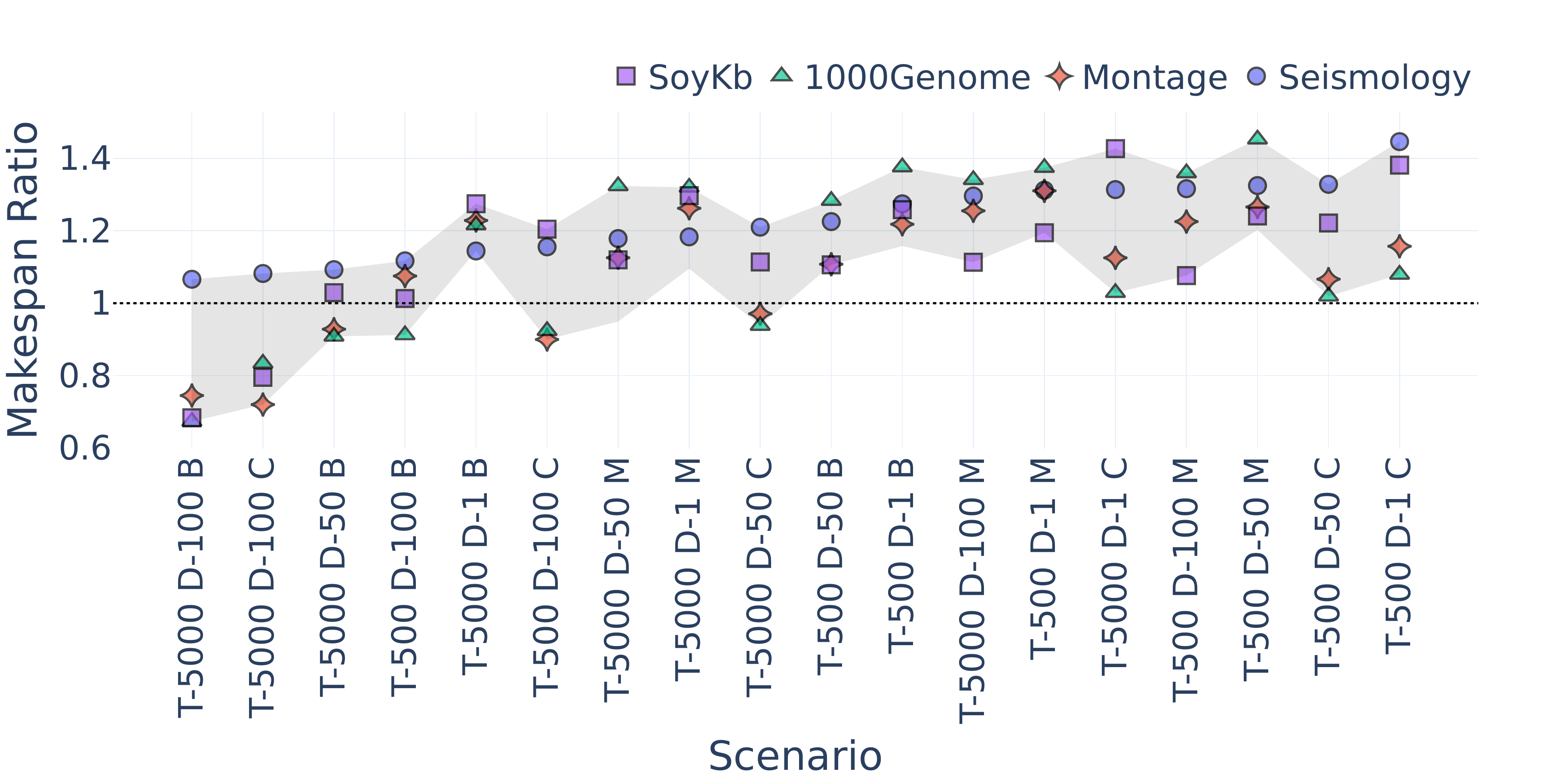}
  \vspace{-20pt}
  \caption{Makespan ratio between workflow executions on 4-node Cascadelake and Skylake platforms. The horizontal axis shows experimental scenarios sorted by increasing Seismology makespan ratios. (T: number of tasks; D: data footprint in GB; C: cpu-bound; M: memory-bound; B: balanced.) Values above (resp. below) $y=1$ correspond to cases in which the Cascadelake execution is faster (resp. slower) than the Skylake execution.}
  \label{fig:makespan_ratio}
  \vspace{-10pt}
\end{figure}

To further motivate the need for workflow benchmarks, we present 
results obtained from the execution of the benchmarks proposed in this work
on two small 4-node (48 cores per ndoe) platforms with 2.6GHz Skylake and 
2.8GHz Cascadelake processors, provided by Chameleon 
Cloud~\cite{keahey2020lessons}. 
Benchmarks are executed using the Pegasus workflow  
system~\cite{deelman-fgcs-2015} and configured for 18 different benchmark 
scenarios. In all scenarios the same amount of compute work is performed, 
but using two different numbers of workflow tasks (500 and 5,000), three 
different total amounts of data to read/write from disk 
(1~GB, 50~GB, and 100~GB), and three different ratios of compute  
to memory operations performed by the workflow tasks (cpu-bound, 
memory-bound, balanced). For each scenario we generated benchmarks 
for workflow configurations that are representative of four different 
scientific workflow application domains (two from bioinformatics, one for 
astronomy, and one for seismology). All details regarding benchmark
generation and configuration are given in 
Section~\ref{sec:approach}. Figure~\ref{fig:makespan_ratio} shows the 
ratio between execution times (or \emph{makespans}) obtained on the Cascadelake 
and the Skylake nodes.

Two key observations can be made from these results. First, results differ 
significantly across workflow configurations, as seen in the width of the 
envelope. Second, trends are difficult to explain. For instance, considering 
the SoyKb and 1000Genome data points, we see that for many scenarios they are 
close to each other, for many scenarios the SoyKb data point is well above 
the 1000Genome data point, and for many other scenarios the situation is 
reversed. Overall, we find that it is difficult to explain, let alone 
predict, workflow (relative) makepans based on platform and workflow 
configurations. Another example of this difficulty is the fact that Skylake 
leads to faster executions for 13 of the 72 benchmark executions. This is 
because these particular Skylake nodes happen to have higher bandwidth disks 
than the Cascadelake nodes. However, for some high-data scenarios (e.g., 
scenarios T-5000-D-100-M  and T-500-D-100-M) Cascadelake executions are 
significantly faster. Furthermore, for the Seismology workflow configuration, 
Cascadelake is always preferable even for high-data scenarios.  Workflow 
performance being difficult to predict is one of the motivations for developing 
workflow benchmarks.

\subsection{Contributions}

In the workflows community, most researchers and practitioners have resorted
to using workflow instances from real-world applications as benchmarks, 
sometimes including these instances as part of benchmark 
suites~\cite{bader2021tarema, capella2017lessons, yang2020edgebench, 
larsonneur2018evaluating}. One drawback is that the obtained results are not 
generalizable, especially because specific workflow instances are not 
configurable and thus may not expose all relevant performance behaviors or 
bottlenecks.  Another drawback, is that executing these benchmarks requires 
installing many scientific software dependencies (since the benchmark code is 
actual application code) and scientific datasets. ``Application skeletons'' 
have been developed that are representative of commonly used workflow 
patterns and can be composed to generate synthetic workflow 
specifications~\cite{katz2016application, coleman2021escience}.  These 
works provide some basis for constructing task-dependency structures 
in workflow benchmarks (to this end this work builds 
on~\cite{coleman2021escience}), but they do not provide fully-specified, let alone executable, benchmarks.


The key insight in this work is that it is possible to automate the
generation of representative workflow benchmarks that can be executed on
real platforms. The main contribution is an approach that implements this 
automation and has the following capabilities: (i)~configurable to be
representative of a wide range of performance characteristics and structures; 
(ii)~instantiable to be representative of the performance characteristics and 
structures of real-world workflow applications; (iii)~automatically 
translatable into executable benchmarks for execution with arbitrary workflow 
systems.  This approach not only generates realistic workflow tasks with 
arbitrary I/O, CPU, and memory demands (i.e., so as to enable weak and 
strong scaling experiments), but also realistic workflow task graphs that are 
based on those of real-world workflow applications, and are agnostic to the 
workflows management system and independent underlying platform. 

The experimental evaluation of our proposed approach is twofold.  First, we
assess the ability of our generated workflow benchmarks to mimic the
performance characteristics of production workflow applications. We do so
by demonstrating that I/O, CPU, and memory utilization for the generated
workflow benchmark tasks corresponds to the performance characteristics of
tasks in real-world workflows, and that, as a result, workflow benchmark 
executions have temporal execution patterns similar to that of real-world 
workflows. Second, we execute a set of workflow benchmarks generated using 
our approach on the Summit leadership-class computing system and compare 
measured performance to that derived from analytical performance models. 

The benefits of these benchmarks are manyfold. Scientists can compare the
characteristics and performance of their workflows to reference benchmark 
implementations; Workflow systems developers can leverage these benchmarks
as part of their continuous integration processes; Computing facilities 
can assess the performance of their systems beyond the traditional HPC 
benchmark implementations; Workflow practitioners can use these
benchmarks to perform fair comparison of competing workflow 
systems.


\section{Related Work}
\label{sec:relatedwork}

The field of HPC has seen the development of many benchmark 
suites~\cite{bailey1991parallel, dongarra2003linpack,hpcchallenge,mantevo, 
luszczek2005introduction, chunduri2019gpcnet, jiang2018hpc, parasyris2020hpc, 
kudo2020prompt, marjanovic2014performance, eigenmann1996benchmarking}. 
For instance, SPEC combines knowledge of performance evaluation with the resources to 
maintain a benchmarking effort by bringing together benchmarking and market 
experts, and customers needs in HPC. In 1994, SPEC's HPG 
emerged to extend the evaluation activities by establishing and 
maintaining a benchmark suite representative of real-world HPC 
applications~\cite{eigenmann1996benchmarking}. More recently, the 
\emph{SPEChpc 2021} suite provides a group of strong-scaled application-based 
benchmarks including metrics per workload size. Although easily 
reproducible, its design limits limit its broadly applicability~\cite{li2022spechpc}.
HPC settings have been 
historically structured around relatively stable technologies and practices 
(e.g., monolithic parallel programs applications that use MPI). Recent 
work~\cite{slaughter2020task} has proposed separating the system-specific 
implementation from the specification of the benchmarks, so as to target 
different runtime systems. This is also the philosophy adopted in this work 
and our benchmarks could easily be implemented within the framework 
in~\cite{slaughter2020task}, which currently does not include 
workflow-specific benchmarks. 


Some researchers have investigated the automatic generation of 
representative benchmarks. For instance, Logan et al.~\cite{logan2012} 
leverage the notion of skeletons to study the I/O behaviors of real-world 
applications. Their approach consists in suppressing computational parts of 
parallel applications, so that only communication and I/O operations remain. 
Users can then run the resulting benchmarks, which exhibit the complex I/O 
and communication patterns of real-world applications, without having to 
experience long execution times. Similarly, Hao et al.~\cite{hao2019automatic} 
leverage execution traces from real-world parallel applications to 
automatically generate synthetic MPI programs that mimic the I/O behaviors 
of these applications without having to execute their computational segments. 

In this work, we focus on scientific workflow applications. Some studies 
have proposed to use particular domain-specific workflows as 
benchmarks~\cite{bader2021tarema, capella2017lessons,
yang2020edgebench, larsonneur2018evaluating, krishnan2021benchmarking}.
For instance, Krishnan et al.~\cite{krishnan2021benchmarking} propose a
benchmark for complex clinical diagnostic pipelines, in which a particular
configuration of a production pipeline is used as a benchmark.
Although these benchmarks are by definition representative of a real-world
application, they are limited to particular scientific domains and application 
configurations. To address this limitation, Katz et 
al.~\cite{katz2016application} and Coleman et
al.~\cite{coleman2021escience} have proposed approaches for generating
synthetic workflow configurations based on representative commonly used
workflow patterns. The limitations are that these works only generate
abstract specifications of workflow task graphs, which is only one of the
required components of an executable workflow benchmark. To the best of our
knowledge, this study is the first to propose a generic workflow benchmark
generation method that makes it possible to generate executable workflow 
benchmarks that can be configured by the users to be representative of a 
wide range of relevant scientific workflow configurations.

\section{Approach}
\label{sec:approach}

Developing a workflow benchmark requires developing (i)~representative
benchmarks of workflow tasks and (ii)~representative benchmarks of workflows
that consist of multiple tasks with data dependencies. We discuss our
approach for each of the above in the next two sections.

\subsection{Developing Representative Workflow Task Benchmarks}
\label{sec:approach:individual}

Workflow tasks have different characteristics in terms 
of compute-, memory-, and I/O-intensiveness~\cite{ramakrishnan2008survey}, 
which impact workflow performance differently on different architectures.
Consequently, a workflow benchmark generation tool should be configurable,
by the user, so that generated benchmark workflow tasks can exhibit
arbitrary such characteristics. 

We have developed a generic benchmark (implemented in Python) that, based
on user-provided parameters, launches instances of different I/O-, CPU-, 
and/or memory-intensive operations. The benchmark executions proceeds in three
consecutive phases\footnote{We do not use stress test tools such as 
\texttt{stress-ng} (e.g., using \texttt{--vm-bytes} or \texttt{--vm-keep} for 
creating memory pressure, or \texttt{--hdd} or \texttt{--hdd-bytes} for 
performing I/O operations) as it does not generate a precise amount of memory 
operations or actual files that could be used downstream in the workflow.}:

\begin{compactenum}
    \item [\#1] {\bf Read input from disk:} Given a binary file, this phase of 
    the benchmark simply opens the file with the ``\texttt{rb}'' option and 
    calls \texttt{file.readlines()} to read the file content from disk, in 
    a single thread. 
  
    \item [\#2] {\bf Compute:} This phase is configured by a number of cores 
    ($n$), a total amount of CPU work ($cpuwork$) 
    to perform, a total amount of memory work to perform ($memwork$), and the 
    fraction of the computation's instructions that correspond to non-memory 
    operations ($f$), which, for now, must be a multiple of $0.1$. 
    This phase starts $n$ groups of 10 threads, where threads in the same group 
    are pinned to the same CPU core (using \texttt{set\_affinity}). 
    Within each group, $10 \times (1-f)$ threads run a memory-intensive 
    executable (compiled C++) that computes random accesses to positions in an 
    array in which one unit is added to each position up to the total amount of 
    memory work ($memwork$) has been performed; and $10 \times f$ threads run a 
    CPU-intensive executable (compiled C++) that calculates an increasingly 
    precise value of $\pi$ up until the specified total amount of computation 
    ($cpuwork$) has been performed. In this manner, our benchmark uses both CPU 
    and memory resources, and parameter $f$ defines the relative use of these 
    resources. For both CPU and memory, the threads are instances of python's 
    \texttt{subprocess} calling the benchmarks executable.
    
    \item [\#3] {\bf Write output to disk:} Given a number of bytes, this phase 
    simply opens an empty binary file with the ``\texttt{wb}'' option and calls 
    \texttt{file.write()} to write random bytes to disk in a single thread.
\end{compactenum}

This above approach is relatively simple and makes several assumptions that
do not necessarily hold true for real-world workflow tasks. For instance,
I/O operations could overlap with computation, and there could be many I/O
and compute phases.  Furthermore, our implementation of the compute
phase (phase~\#2) on the CPU uses multiple threads that can have complex 
interference in terms of resource usage (e.g., cache vs. main memory use).  
Furthermore, due to our use of 10 threads per core, there is context-switching 
overhead that likely does not occur with real-world workflow tasks. Finally, 
due to our use of only 10 threads, $f$ can only take discrete values (multiples 
of 0.1), which does not make it possible to capture arbitrary non-memory/memory 
operation mixes.  Nevertheless, we claim that this approach makes it possible to 
instantiate benchmarks that are representative of real-world workflow tasks. 
We verify this claim in Section~\ref{sec:validation:individual}.

\subsection{Developing Representative Workflow Benchmarks}
\label{sec:approach:workflow}

Now that we have an approach for developing benchmarks of workflow tasks,
we need an approach for producing a benchmark of an entire workflow of
these tasks. To this end, we rely on the recently developed
WfChef~\cite{coleman2021escience} open source tool. Given a set of real
workflow instances for a particular scientific application, WfChef analyzes the
task graphs in these instances to produce a ``workflow recipe", i.e., data 
structures and code that describes discovered task-dependency patterns. A
workflow recipe can then be used to generate synthetic workflow
task graphs with (almost) arbitrary numbers of tasks. The results
in~\cite{coleman2021escience} show that WfChef is able to produce synthetic
workflow task graphs with realistic structures that are representative of 
those found in real-world workflows. 

Given the above, we have developed a workflow benchmark generator 
that takes as input a desired number of tasks and a WfChef workflow
recipe.  In~\cite{coleman2021escience} the authors have generated workflow recipes for many scientific applications. We use these same
recipes for a subset of these applications
for our experimental evaluations in 
Sections~\ref{sec:validation} and~\ref{sec:results}.  Our workflow
benchmark generator first invokes the WfChef recipe to generate a task
graph.  Once the task graph has been generated, each task is set to be an
instance of the workflow task benchmark described in the previous section.
For each task, the user can specify values for the parameters of the workflow task
benchmark described in the previous section that pertain to the computation
($n$, $cpuwork$, $memwork$, $f$). The user can specify individual data 
volumes for each task in a way that is coherent with respect to task data 
dependencies. Alternatively, the user can specify a total data footprint, i.e., 
the sum of the sizes in bytes of all data files read/written by workflow tasks, 
in which case uniform I/O volumes are computed for each workflow task
benchmark.

\begin{figure}[!t]
  \centering
  \includegraphics[width=.75\linewidth]{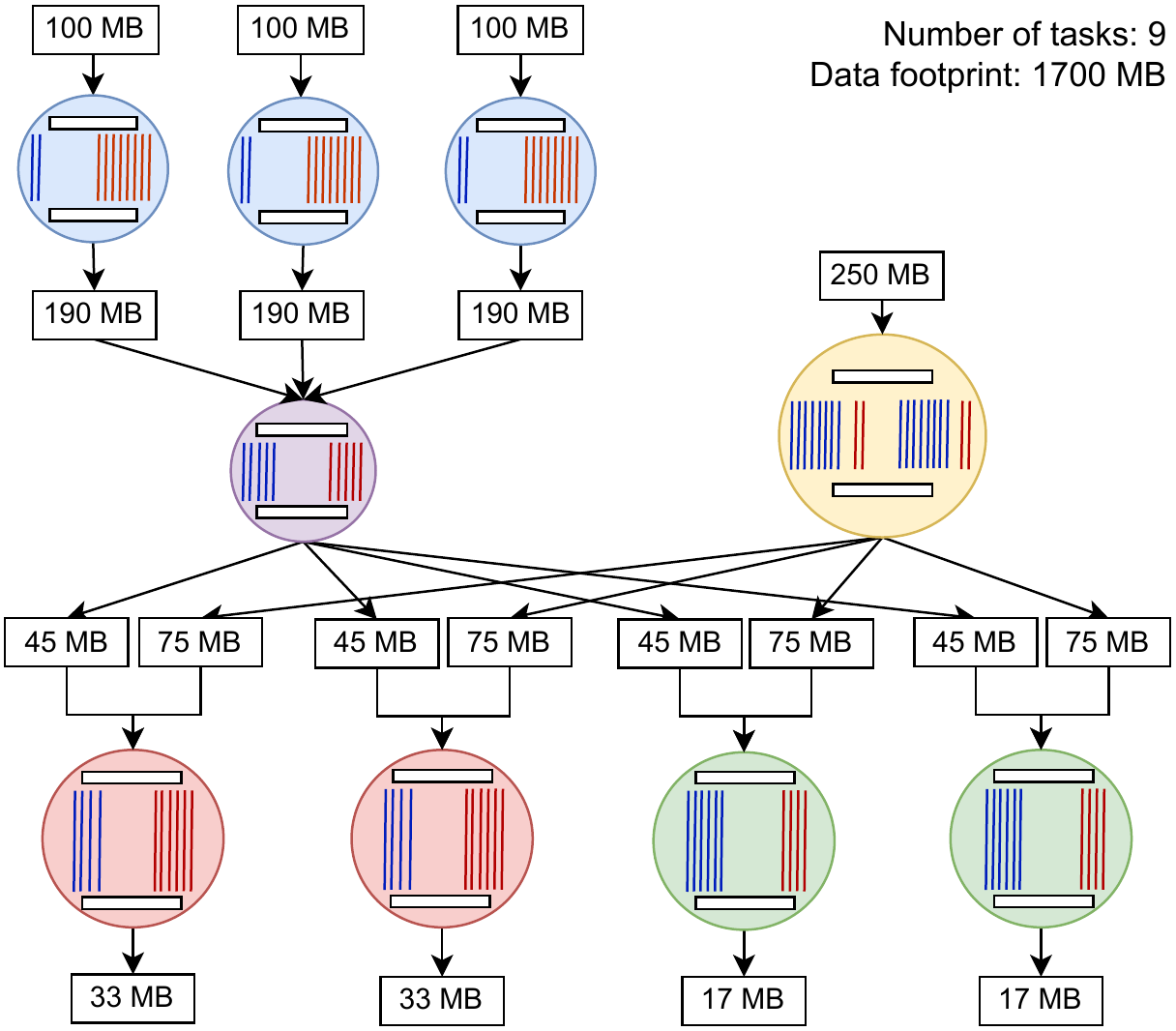}
  \vspace{-8pt}
  \caption{Example of a generated workflow benchmark. For each task,
           represented as a circle, inner rectangles represent I/O operations 
           performed in a single thread (phase~\#1, read input from disk, 
           and phase~\#3, write output to disk); blue/red lines represent
           the CPU/memory threads respectively (phase~\#2). Outer white rectangles
           denote data files and show their respective sizes. Task data dependencies are depicted as directed edges.}
  \label{fig:benchmark_overview}
  \vspace{-10pt}
\end{figure}

Figure~\ref{fig:benchmark_overview} illustrates a CPU-only benchmark 
instantiation with nine tasks (each task use $n=1$ core but for the 
yellow task, which uses $n=2$ cores). Each task has a different mix of cpu- and 
memory-intensive threads (shown as blue and red lines), due to different tasks 
having different values of the $f$ parameter. (The figure does not depict that 
different tasks may have different $cpuwork$ values.) The benchmark also includes 
19 data files, for a total data footprint of 1,700~MB, which are input/output of 
various tasks so as to create the particular task-dependency structure shown by 
the edges in the figure.

Our generator returns a JSON object that fully describes the workflow
benchmark in terms of tasks, task performance characteristics, task input
and output files, and task data dependencies. This JSON object, along with
the workflow task benchmark implementation described in the previous
section, can be used to generate an executable workflow. For instance,
for the experiments in Sections~\ref{sec:validation:full} 
and~\ref{sec:results}, we implemented a translator that translates the JSON 
object into programs for executing workflow benchmarks using the
Pegasus~\cite{deelman-fgcs-2015} and Swift/T~\cite{wozniak2013swift} workflow 
systems, respectively\footnote{Translators can be written 
for most DAG-based workflow 
systems~\cite{workflow-systems}.}.

\section{Validating the Accuracy of Generated Workflow Benchmarks}
\label{sec:validation}

In this section, we present experimental evaluation results to quantify 
the soundness of our benchmarking approach. These experiments are performed 
on four different kinds of (dedicated) compute nodes
as listed in Table~\ref{tab:platforms}. 
Several Haswell, Skylake, and Cascadelake nodes are provided 
by the Chameleon Cloud~\cite{keahey2020lessons}. One older nehalem node 
is a bare-metal server hosted at one of our institutions. 

\begin{table}[!t]
  \scriptsize
  \centering
  \setlength{\tabcolsep}{4.5pt}
  \caption{Compute node hardware configurations used to perform validation experiments.}
  \vspace{-8pt}
  \label{tab:platforms}
  \begin{tabular}{llrrr}
    \toprule
    \multirow{2}{*}{Family Name} & \multirow{2}{*}{Processor} & \multirow{2}{*}{\#cores} & RAM & LLC \\
    & & & (GB) & (MB) \\
    \midrule
    nehalem     &  Intel Xeon E5530 CPU @ 2.4 GHz & 16 & 24 & 8\\
    Haswell     &  Intel Xeon E5-2670 CPU @ 2.3 GHz & 48 & 128 & 32\\
    Skylake     &  Intel Xeon Gold 6126 CPU @ 2.6 GHz & 48 & 196 & 19\\
    Cascadelake &  Intel Xeon Gold 6242 CPU @ 2.8 GHz & 64 & 196 & 22\\
    \bottomrule
  \end{tabular}
  \vspace{-10pt}
\end{table}

\subsection{Workflow Tasks}
\label{sec:validation:individual}

To evaluate the benchmarking approach described in 
Section~\ref{sec:approach:individual}, we consider four different tasks
from the Montage astronomy workflow~\cite{deelman-fgcs-2015} and three different 
tasks from the 1000Genome bioinformatics 
workflow~\cite{ferreiradasilva-fgcs-2019}, which altogether correspond to seven 
different compiled executables. Our objective is to verify whether it is possible 
to configure our workflow task benchmark so that its performance behavior is
similar to that of each of these workflow tasks. That is, the benchmark's
execution time should track that of the workflow task on a particular
compute node for various load conditions.

\pp{Experimental Goals}
Recall from Section~\ref{sec:approach:individual} that our CPU-only 
benchmark is
configured by a number of cores ($n$), an amount of CPU
work ($cpuwork$), an amount of memory work ($memwork$), and a fraction of the 
executed CPU instructions that are non-memory operations ($f$). 
In this section, 
we ignore the I/O portion of the
benchmark, since it can simply be configured based on the I/O volumes
observed in the execution of the real workflow task. More challenging is
the computational portion of the benchmark, and in particular the configuration
parameter $f$. We want to answer two questions: (i)~is it possible to pick
a good value for $f$? and (ii)~does picking a good value matter?

To answer these two questions, we perform the following experiments. Given
a workflow task's execution on a core of a particular compute node, we
instantiate our benchmark (i.e., we pick $cpuwork$ and $memwork$ values) for each value of $f$ (i.e., 0.1, 0.2, \ldots,
0.9) so that its execution time is equal to that of the real workflow task
on that compute node. Then, we execute the real workflow task and all these
benchmark instantiations on the same compute node again, but with
additional external memory load.  The goal is to have the compute node
exhibit a different relative performance of CPU and memory.  As a result,
the executions of the workflow task and of the instantiated benchmarks are
slowed down, and we wish to confirm that: (i)~there exists a value of $f$
that makes our benchmark's execution time track the execution time of the real
workflow task; and that (ii)~this value of $f$ remains the same across
different compute nodes with different architectures and regardless of the 
external load. If both hypotheses are confirmed, then parameter $f$ 
is useful and it is possible to pick a good value for it.


\pp{Experimental Methodology}
On a compute node, we execute a real workflow task and measure its execution 
time, $T$.
For each possible value of $f$, we then search for the $cpuwork$ and $memwork$
values that make the benchmark run (approximately) in time $T$.  We perform
this search greedily as follows. We start with two guesses for $cpuwork$ and
$memwork$, run the benchmark, and measure the completion times of the
cpu-intensive threads, $T_{cpu}$, and of the memory-intensive threads,
$T_{mem}$.  We then adjust the amounts of work as $cpuwork = cpuwork \times
T/T_{cpu}$ and $memwork = memwork \times T/T_{mem}$. We repeat this process
until $|T_{cpu}-T|/T < 5\%$ and $|T_{mem}-T|/T <5\%$, which converges after
only a few iterations in all our experiments.  We then re-run the workflow
task and the instantiated benchmark on that same compute node, but on which we
have now introduced extra memory load. This load is introduced by running
\texttt{stress-ng} instances on other cores of the compute nodes. More
precisely, we run one instance of \texttt{stress-ng} on $n/2$ other cores ($n$
is the total number of cores on the compute node), which causes sufficient
memory load to impact the execution of the workflow task and of our
benchmark without leading to execution times that would be impractically high.  
For each value of $f$, we can then compare the benchmark's execution time and
the workflow task's execution time. We perform these experiments on all
the compute node hardware configurations listed in Table~\ref{tab:platforms}.

\begin{figure*}[!t]
  \centering
  \includegraphics[width=.75\linewidth]{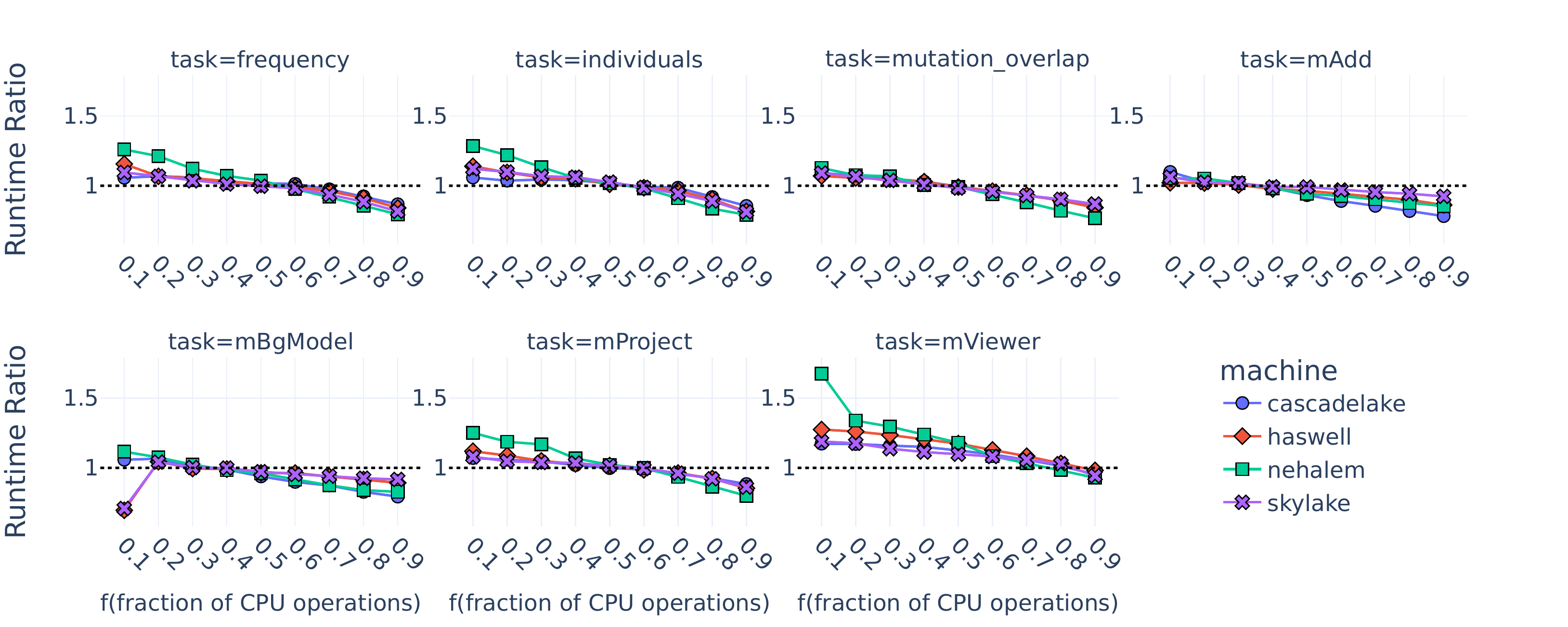}
  \vspace{-10pt}
  \caption{Ratio between the benchmark's execution time and that of the 
           workflow task for different fractions of the computation's 
           instructions that correspond to CPU operations ($f$).}
  \label{fig:validation}
  \vspace{-10pt}
\end{figure*}

\pp{Results}
Experimental results are shown in Figure~\ref{fig:validation}. Each plot is
for different Montage and 1000Genome tasks and shows the ratio between the
benchmark's execution time and that of the workflow task (vertical axis)
vs. $f$ (horizontal axis). Values above (resp. below) $1.0$ correspond
to cases in which the benchmark execution is longer (resp. shorter) than
that of the workflow task. Each curve represents a different compute
node.  We can make several observations from these results. 

First, the
execution time ratio varies based on $f$, in the range 0.7-1.68. That is,
for a ``bad" $f$ value, the benchmark's execution time can be up to 30\%
(reps. 68\%) faster (resp. slower) than that of the workflow task.  Second,
we find that for each compute node architecture there is a ``good" $f$ value that
leads to a ratio close to $1.0$, and that this value is different for each
workflow task.  This means that it is possible to find a good $f$ value for
each workflow task, and that having $f$ be constrained to be a multiple of
$0.1$ (so that each benchmark uses only 10 threads), is sufficient to
obtain good results.  Unsurprisingly, we find that different workflow tasks
have different performance characteristics in terms of CPU and memory usage
(best $f$ values for our seven tasks vary between 0.4 and 0.8).  Third, and
most importantly, we find that for a given workflow task the same $f$ value
is consistently best (or close to best) across all compute node
architectures. That is, in each plot, all curves intersect the $y=1$ line
for almost the same $x$ values. This observation validates that our benchmark 
can be instantiated to capture a workflow task's fundamental performance
characteristics and thus have execution behavior that is representative of
that of the workflow task across different architectures.

An interesting question is: Given a workflow task, how can we determine the 
best $f$ value for instantiating a representative benchmark of that task? 
One option is to run the above experiments so as to determine the best $f$ 
value empirically. Given the results in Figure~\ref{fig:validation}, running 
these experiments on a single compute node should suffice. Alternately, one 
could consider profiling the task's execution on a particular compute node 
to discover its relative CPU and memory usage. This can be done for instance 
using the \texttt{perf} Linux tool to measure the total number of hardware 
instructions executed and how many of these are memory load/store instructions. 
To evaluate the effectiveness of this approach, we profiled the execution of 
our seven workflow tasks on a nehalem compute node. Table~\ref{tab:perf} shows 
how the $f$ values determined based on profiling with \texttt{perf} compared 
to those based on the experimental results in Figure~\ref{fig:validation}.  
We show profiling-based values rounded off to the nearest multiple of $0.1$, 
which would presumably be used by a user as the $f$ value provided to our 
benchmark, but also, in parentheses, the values before rounding off.

\begin{table}[!t]
  \setlength{\tabcolsep}{12pt}
  \scriptsize
  \centering
  \caption{Estimated values for the fractions of the computation's
           instructions that correspond to CPU operations ($f$) obtained 
           with the \texttt{perf} Linux tool and our empirical experiments.}
  \vspace{-5pt}
  \begin{tabular}{lrr}
    \toprule
    \multirow{2}{*}{Task} & \multicolumn{2}{c}{values for $f$} \\
    &  \texttt{perf} &  empirical \\
    \midrule  
      mAdd              &   0.5 (0.45) &        0.4 \\
      mViewer           &   0.4 (0.38) &        0.8 \\
      mBgModel          &   0.5 (0.54) &        0.4 \\
      mProject          &   0.6 (0.55) &        0.6 \\
      individuals       &   0.6 (0.57) &        0.5 \\
      mutation\_overlap &   0.6 (0.62) &        0.4 \\
      frequency         &   0.5 (0.53) &        0.6 \\
    \bottomrule
  \end{tabular}
  \label{tab:perf}
  \vspace{-10pt}
\end{table}

We find that using the profiling approach produces the empirically best $f$
value, once rounded off, for one of our seven workflow tasks, and is within 0.1
of the best $f$ value for all but two of these seven tasks (for which it is 0.2
and 0.4 away). It is expected that the profiling approach cannot always
produce the best $f$ value (e.g., because our memory-intensive threads also 
use the CPU in addition to performing memory operations).  But it may still be 
attractive since, in most cases, it makes it possible to find a reasonable $f$ 
value by running the workflow task only once.  This said, although a $0.1$ 
deviation from the best $f$ values may seem low, this deviation can have a 
large impact on benchmark accuracy for scenarios in which multiple workflow
tasks run concurrently on the same compute node.

Overall, we conclude that our workflow task benchmark, albeit simple, can
be configured to be representative of the fundamental performance 
characteristics of real workflow tasks.

\subsection{Workflows}
\label{sec:validation:full}

In this section, we verify the claim that our approach can generate
benchmarks representative of full workflow applications. 
For each Montage and 1000Genome task, 
we instantiate a representative workflow task 
benchmark -- I/O volumes are based on the actual input/output file sizes
as specified in the workflows. We empirically 
determine the other benchmark configuration parameters that lead to 
benchmark task execution times that are close to actual task execution times 
and remain so under different load conditions.
We then generate a representative task data-dependency structure
using the approach described in Section~\ref{sec:approach:workflow}.  
This results in a benchmark instantiation, the specification of which is 
produced as a JSON file. 
From this JSON file, we then automatically generate a Python program for 
executing the workflow. 
This allows us to execute the benchmark using Pegasus 
in exactly the same way in which Montage and 1000Genome workflows are executed 
in production. Consequently, we can perform sound comparisons 
between benchmark executions and Montage and 1000Genome executions, where 
both kinds of executions use the exact same software stack. 

\begin{table}[!t]
  \centering
  \scriptsize
  \setlength{\tabcolsep}{5pt}
    \caption{Comparison between the real-world Montage workflow application 
             and the generated workflow benchmark.}
    \label{table:montage-runs}
    \vspace{-5pt}
    \begin{tabular}{lrrrrr}
      \toprule  
      \multirow{2}{*}{$degree$} & \multicolumn{2}{c}{\#tasks} & \multicolumn{2}{c}{makespan (sec)} & \multicolumn{1}{c}{makespan. time}\\
      & Montage  & Benchmark &  Montage & Benchmark & \% difference\\
      \midrule
      1.0 & 157  & 157  & 709   & 822   & 13.74 \%\\
      1.5 & 707  & 717  & 1582  & 1756  & 10.99 \%\\
      2.0 & 2149 & 2130 & 2017  & 2220  & 10.06 \%\\
      2.5 & 5155 & 5155 & 10015 & 9957  & -0.58 \%\\
      \bottomrule
    \end{tabular}
    \vspace{-10pt}
\end{table}

\pp{Montage workflows}
Table~\ref{table:montage-runs} shows benchmark and Montage results for four
different configurations of the Montage application. Specifically, each
configuration is for a different value of the $degree$ input parameter to
Montage, which has a large impact on the size of the workflow, as seen in
the number of tasks shown in the second column of the table.  Note that
our benchmark does not necessarily contain that exact same number of tasks.
While this may seem surprising, it is an artifact of the WfChef synthetic
workflow generation approach described in
Section~\ref{sec:approach:workflow}. Recall that WfChef generates workflow
structures based on real workflow instances for a particular application.
To do so, it ``mines" these instances to discover repeated sub-structures,
and then replicates these sub-structures to generate a workflow of a
particular size. As a result, it cannot generate synthetic workflows for
all arbitrary numbers of tasks. In our case, for instance, for $degree =
1.5$, although we invoke WfChef asking it to generate a workflow with 707
tasks, it returns a workflow with 10 more tasks.  For $degree = 2.0$, it
generates a workflow with 19 fewer tasks than desired. These differences in
numbers of tasks have an impact on overall execution time,  but, given the
current WfChef design and implementation, they are the price to pay for having
realistic task-dependency structures in our benchmarks.

All executions are performed on an HTCondor~\cite{thain2005distributed} 
cluster composed of four Haswell compute nodes using Pegasus. During these executions, many workflow/benchmark tasks are executed
concurrently on different cores of the same compute node, and thus interfere
with each other for memory and I/O operations. (Each workflow/benchmark 
application is executed separately, thus only tasks from the same 
workflow/benchmark are executed concurrently.)
Makespans are reported in the fourth and fifth 
columns of Table~\ref{table:montage-runs},
with the percentage difference, relative to the execution of the
real Montage workflow, shown in the sixth column. These results show that
benchmark executions have makespans within at most 14\% of the
makespans of the real workflow. For the largest workflow
configurations ($degree = 2.5$), the makespans are the closest, with
less than 1\% difference. 

\begin{figure}[!t]
  \centering
  \includegraphics[width=\linewidth]{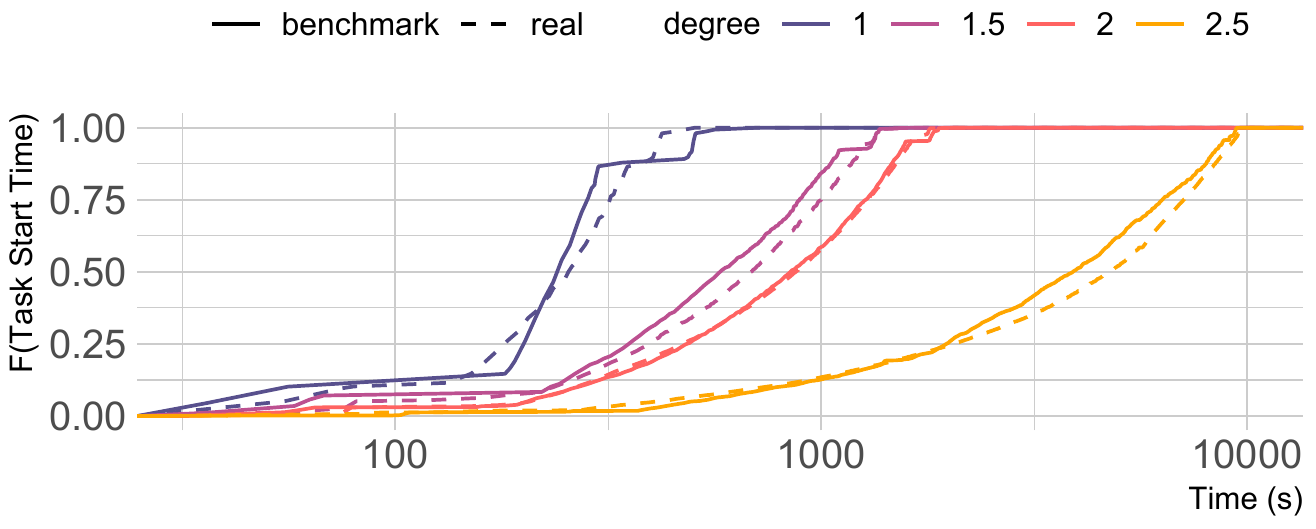}
  \vspace{-20pt}
  \caption{Empirical cumulative distribution function of task start times
           for sample real-world and benchmark workflows for the Montage 
           application.}
  \label{fig:montage-runs}
\end{figure}

The results in Table~\ref{table:montage-runs} show that benchmark makespans
are close to real makespans, even though our task benchmarking
approach makes several simplifying assumptions that do not necessarily hold
for real-world tasks (see discussion in
Section~\ref{sec:approach:individual}) and our benchmark workflow does not
have the exact same task-dependencies and not always the same number of tasks
as the real workflow.  However, the makespan is only one metric
for quantifying the discrepancy between benchmark and real workflow
executions. One may wonder whether the temporal structures of the
executions are also similar.  Figure~\ref{fig:montage-runs} shows the empirical
cumulative distributed functions (ECDF) of task start times for each
real Montage workflow (dashed lines) and its benchmark counterpart (solid
lines). We observe the ECDF for the benchmark execution is close to
that for the real execution, with similar overall shape and
inflection points.

\begin{table}[!t]
  \centering
  \scriptsize
  \setlength{\tabcolsep}{4.5pt}
  \caption{Comparison between the real-world 1000Genome workflow application 
           and the calibrated workflow benchmark.}
  \label{table:genome-runs}
  \begin{tabular}{crrrrr}
    \toprule  
    \multirow{2}{*}{\#ch}  &  \multicolumn{2}{c}{\#tasks} & \multicolumn{2}{c}{makespan (sec)} & \multicolumn{1}{c}{makespan}\\
    & 1000Genome  & Benchmark &  1000Genome & Benchmark & \% difference\\
    \midrule
    1 & 66   & 66   & 971   & 892   & -8.13 \%\\
    2 & 232  & 234  & 2548  & 2531  & -0.66 \%\\
    3 & 648  & 645  & 6321  & 5645  & -10.56 \%\\
    4 & 1548 & 1548 & 13409 & 13553 & 1.07 \%\\
    \bottomrule
  \end{tabular}
  \vspace{-10pt}
\end{table}

\begin{figure}[!t]
    \centering
    \includegraphics[width=\linewidth]{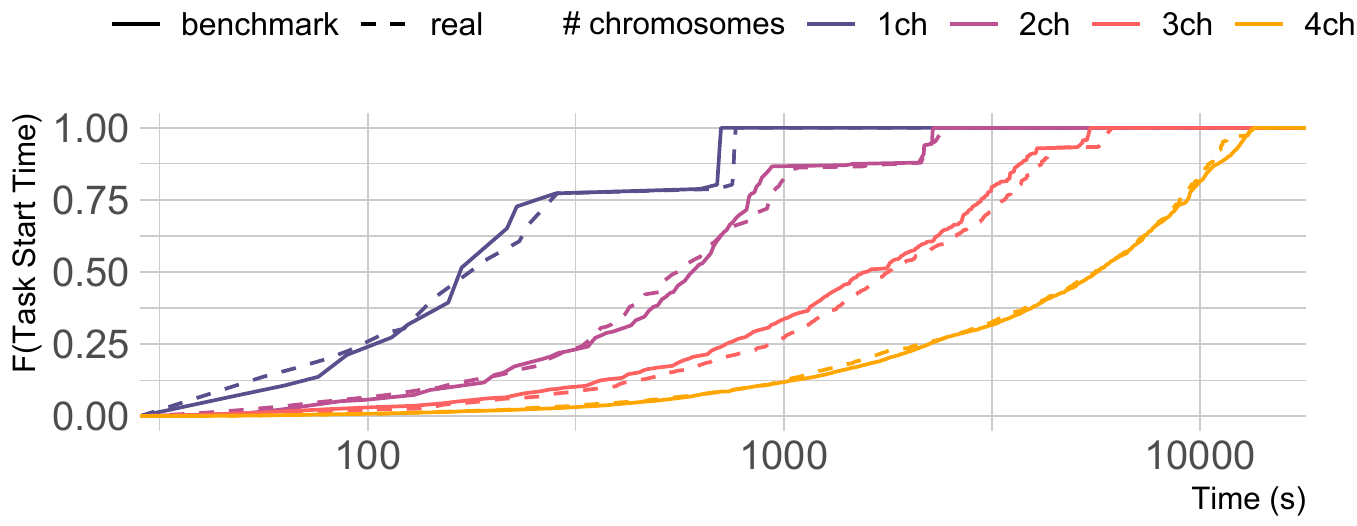}
    \vspace{-20pt}
    \caption{Empirical cumulative distribution function of task start times
           for sample real-world and benchmark workflows for the 1000Genome
           application.}
    \vspace{-10pt}
    \label{fig:genome-runs}
\end{figure}

\pp{1000Genome workflows}
Table~\ref{table:genome-runs} shows benchmark and 1000Genome results for four
different configurations of the 1000Genome application. Specifically, each
configuration is for a different value of the $ch$ (number of chromosomes) 
input parameter to 1000Genome, which has a large impact on the size of the 
workflow, as seen in the numbers of tasks shown in the second column of the 
table. Similarly to Montage, all executions are performed on the HTCondor
cluster using Pegasus. All relative makespan differences are within
11\%, noting that for $ch=2$ and $ch=4$ the difference is about 1\%. 
Figure~\ref{fig:genome-runs} shows the ECDF of task start times. Like for
the Montage results, visual inspection shows that the 
ECDF for the benchmark execution is close to that for the real 
workflow execution, with similar overall shape and inflection points.

Overall, we conclude that our approach makes it possible to generate
workflow benchmark that have structure, performance characteristics, and
execution patterns, that are very similar to that of real-world workflow
applications.

\section{Experimental Evaluation}
\label{sec:results}

In this section, we measure the impact of
different instances of IO- and CPU-intensive operations on workflow applications
that exhibit different workflow patterns. Specifically, we generate CPU-only 
benchmarks for five very distinct workflow applications (Blast, Cycles, 
Epigenomics, Montage, and SoyKB). These applications exhibit a range of 
task-dependency structures (deep fan-out-fan-in tasks pipelines, 
highly-parallel shallow task-graphs, etc.). The goal of these experiments is
to show that it is possible, using our approach, to generate a suite of workflow
benchmarks that make it possible to uncover non-trivial performance behaviors 
that occur on a specific platform using a specific workflow system.


\pp{Experiment Scenario}
For each workflow application, we generate workflow instances composed of 1,000, 
10,000, 50,000, and 100,000 tasks, and for each number of tasks configuration 
we generate instances in which the total workflow data footprint is 100~GB and 
1~TB. 
In total, we generate 40 different workflow configurations for this 
experiment. We run these workflow instances on the ORNL's Summit leadership class
HPC system~\cite{vazhkudai2018design}. Summit is equipped with 4,608 compute 
nodes, in which each is equipped with two IBM POWER9 processors (42 cores), 
six NVIDIA Tesla V100 accelerators each with 96~GiB of HBM2, 512~GB of DDR4 
memory, and connection to a 250~PB GPFS scratch filesystem. Workflow executions 
are performed using the Swift/T~\cite{wozniak2013swift} workflow system. 
We chose Swift/T as its workflows are compiled into MPI programs that are 
optimized for running at scale on HPC clusters. Each workflow is configured to 
use up to 40 CPU cores per compute node (two cores are reserved for management 
operations), and the total number of nodes requested per workflow is computed as 
$(0.1 \times \#tasks)/40$, e.g. a workflow instance composed of 10,000 tasks 
uses 25 nodes (or 1,000 CPU cores), and a 100,000-tasks instance uses 250 nodes 
(or 10,000 CPU cores). All workflow tasks compute the same amount of work 
($cpuwork=500$, $memwork=0$), so comparison across workflow 
configurations is consistent.

\begin{figure}[!t]
    \centering
    \includegraphics[width=\linewidth]{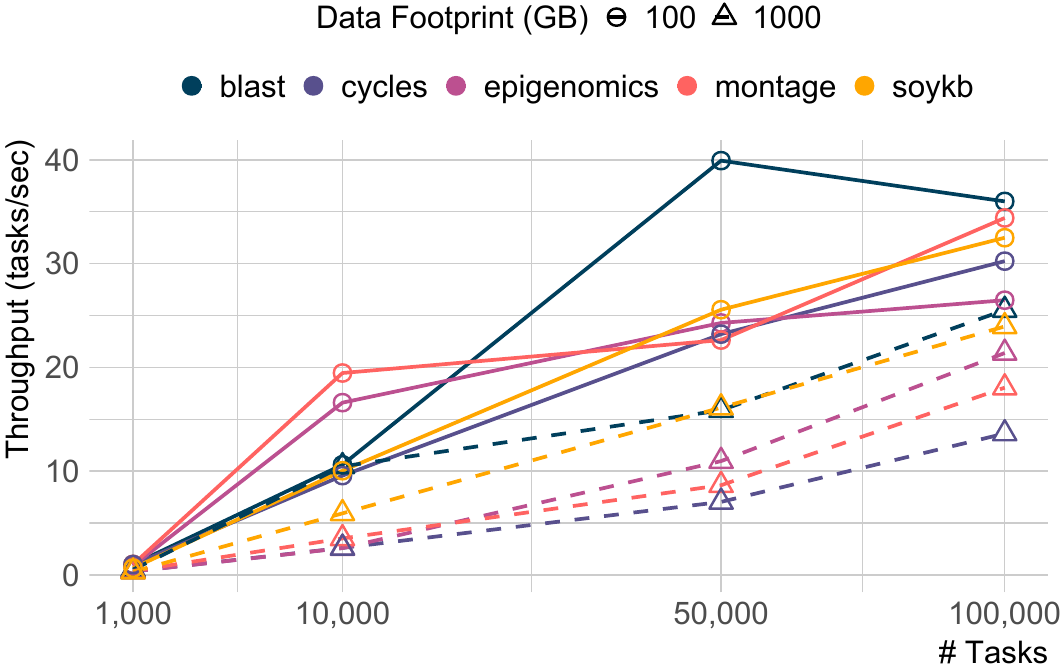}
    \vspace{-20pt}
    \caption{Workflow throughput weak scaling for different number of tasks 
            in the workflow and total number of CPU cores used (i.e., 
            $0.1\times\#tasks$). Runs were performed using ORNL's Summit 
            leadership class HPC system.}
    \vspace{-10pt}
    \label{fig:summit-overall}
\end{figure}

\pp{Results}
Figure~\ref{fig:summit-overall} shows workflow throughput (number of tasks per 
second) for all workflow configurations. As data footprint increases
workflow throughput is impaired by the time spent on I/O operations; and
workflow throughput improves with the size of the workflow (due to increased 
parallelism). In addition to these expected results, these experiments also
show several less expected results. Although workflows of same size perform the 
same total amount of CPU work, I/O load distribution may significantly diverge 
across workflows---the task graph structure may encompass highly parallel batches 
of tasks (e.g., Blast), or sparse sets of specific workflow patterns (e.g., 
Cycles)---and therefore, I/O contention may occur at different times throughout 
the workflow execution. This can have a large impact on workflow performance, 
as seen for instance for the Blast and Cycles workflows: a Blast workflow 
can perform about twice as many tasks per time unit than a Cycles workflow for 
50,000-tasks instances. Conversely, for low data footprint (100~GB) and 
10,000-tasks configurations, Montage and Epigenomics instances outperform the 
other workflow applications by up to a factor 1.9. This is due to their  
task graph structures---they both present a fan-out-fan-in pattern that is 
repeated as the number of tasks in the workflow increases. In a low data 
footprint configuration, fan-in tasks do not experience I/O bottlenecks due 
to the number of files generated by their parent tasks, but as the workflow 
grows in size and data footprint, performance is significantly impacted by 
data read/write operations.

\begin{figure}[!t]
    \centering
    \includegraphics[width=\linewidth]{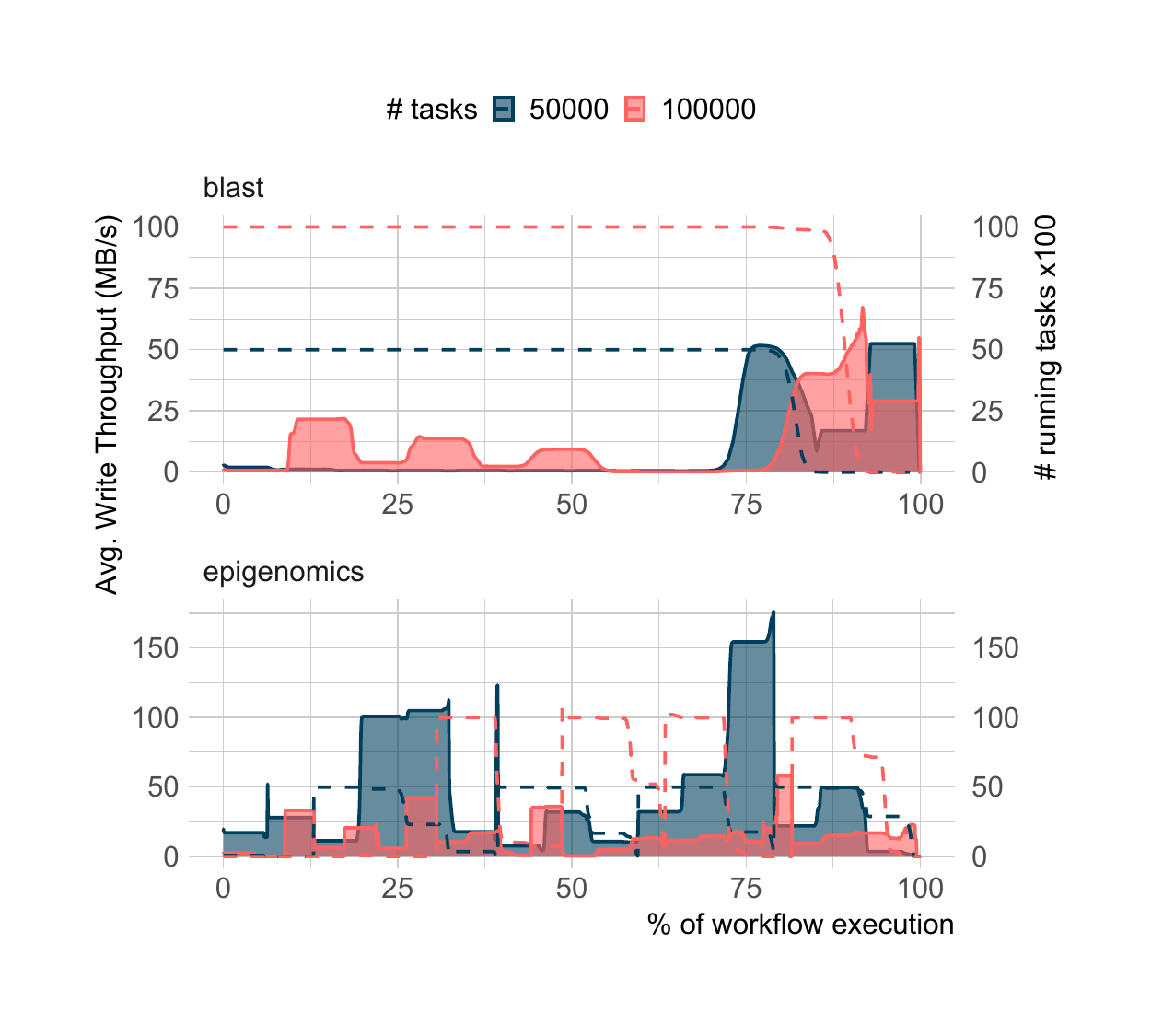}
    \vspace{-20pt}
    \caption{Average write throughout (in MB/s, overlapping area chart) and 
            number of concurrent running tasks (dashed-lines) for 50,000- and 
            100,000-tasks Blast (\emph{top}) and Epigenomics (\emph{bottom}) 
            workflow instances. Each workflow run has a 100GB data footprint.}
    \vspace{-10pt}
    \label{fig:summit-write}
\end{figure}

In Figure~\ref{fig:summit-overall}, both Blast and Epigenomics achieve lower
throughput for 100,000-task instances when compared to 50,000-task instances 
for the low data footprint configuration. An examination of the individual 
execution logs for these runs reveals that there is a considerable difference 
in I/O write throughput between 50,000-task and 100,000-task instances.  
Figure~\ref{fig:summit-write} shows average write throughput measurements 
(overlapping area chart) and number of running tasks (dashed lines) throughout 
the workflow execution for both applications. Overall, write throughput is 
noticeably lower for the large instances (up to a factor 10 for Epigenomics). 
The source of this low performance is twofold. First, individual files have 
relatively small sizes ($\sim$20~MB and $\sim$10~MB for 50,000 and 100,000 tasks, 
respectively), which is known to hinder the ability of the I/O system to reach 
high throughput. Second, the 100,000-task instances operate over twice as many 
files, which may increase contention for the shared file system. The latter is 
evidenced by spikes in write throughput when the number of concurrently running 
tasks decreases.  These observations showcase the fact that understanding, and 
thus modeling, workflow performance accurately is not easy, and observation that 
we further corroborate in the next section.

\section{Usefulness of Benchmarks}
\label{sec:model}

A common use of benchmarks is to estimate overall execution time, or 
\emph{makespan}, for different application and/or platform configurations. In 
Section~\ref{sec:results}, we show benchmark results for the Summit platform. 
One may wonder whether these results could be estimated using simple performance 
models. If the obtained estimates make it possible to compare makespans across 
different configurations reliably, then perhaps workflow benchmarks are not 
useful, at least for the purpose of makespan estimation. In this section, we 
present a few makespan models and assess whether they can be used to estimate 
workflow performance accurately. 

\subsection{Workflow Makespan Models}

Consider a platform that comprises $n$ $p$-core compute nodes. Each node can 
read, resp. write, data to some shared file system with a data rate of 
$bw_{read}$, resp. $bw_{write}$, in bytes/sec. We assume that both these data 
rates are measured by running a simple I/O benchmark (e.g., using the \texttt{dd} 
command). Note that concurrent I/O operations performed by multiple nodes may be 
limited by some overall bottleneck aggregate bandwidth to the shared file system. 
We leave this out of our model because the Summit platform advertises very high 
such aggregate bandwidth. Consider a workflow benchmark that is to be executed on 
this platform, where each task $t$ runs on a single core and has a known work 
$w_t$ in seconds. We assume all $w_t$ values are measured by executing each task 
individually. We have performed the above measurements on the Summit platform 
($w_t$ = 20.62 sec for all workflow benchmark tasks; $bw_{read}$ = 466 MB/sec; 
$bw_{write}$ = 60 MB/sec).

\pp{Macro-task Models}
A simple approach when modeling a complex application, such as a workflow, is 
to view the application as a single ``macro" task that reads input, performs 
computation, and writes output. Given any workflow, we can compute the total 
amount of data read by the workflow tasks in bytes ($data_{read}$), the total 
amount of data written by the workflow tasks in bytes ($data_{write}$), and the 
total amount of sequential work performed by the workflow tasks ($work$). The 
workflow execution time can then be estimated as:
\vspace{-6pt}
\begin{eqnarray*}
makespan &=& data_{read} / (n \times bw_{read}) +\\
& & work / (n\times p) +\\
& & data_{write} / (n \times bw_{write})\;.
\end{eqnarray*}
\noindent
The second term above assumes that the computation can be parallelized perfectly 
across all cores, which is typically not the case due to task dependencies. 
The above model assumes no overlap between computation and I/O, but such overlap 
occurs in workflow executions. A simple model that assumes perfect overlap is as 
follows:
\vspace{-6pt}
\begin{eqnarray*}
makespan &=& \max(work / (n\times p), \\
& & data_{read} / (n \times bw_{read}) +\\ 
& & data_{write} / (n \times bw_{write}))\;.
\end{eqnarray*}

\pp{Per-level Model} 
The Macro-task models ignore the task dependency structure of the 
workflow. To develop a more accurate model that accounts for the workflow 
structure, we consider that the workflow consists of $L$ levels, where level 
$l = 0, \ldots, L-1$ is the set of tasks with a top-level equal to 
$l$. The top-level of a task is defined as the length, in number of edges, of 
the longest path from workflow entry task to that task. Thus, level $0$ consists 
of the workflow's entry tasks. All tasks in the same level can be performed 
concurrently, and when all tasks in level $l$ have been completed it is 
guaranteed that all tasks in level $l+1$ can begin execution. We assume a 
level-by-level execution so that the makespan is estimated as:
\vspace{-6pt}
\begin{equation*}
makespan = \sum\nolimits_{l=0}^{L-1} makespan(l)\;,
\end{equation*}
where $makespan(l)$ denotes the makespan of level $l$.  Note that in practice 
there could be overlap between the execution of consecutive levels. 

A difficulty here is that $makespan(l)$ depends on how the tasks in level $l$ 
are scheduled onto the available cores, and in particular the order in which 
they are scheduled. The scheduling scheme depends on the workflow system, 
and it is unlikely that a user would be able to reverse-engineer this scheme 
precisely. We assume that tasks are sorted in non-increasing order of execution 
time (i.e., I/O time plus compute time) and scheduled in this order, which 
corresponds to a standard list-scheduling approach. Rather than computing a 
precise schedule (which would amount to running a full-fledged simulation), 
we assume that tasks are executed in batches of at most $n\times p$ 
tasks, where all batches, save perhaps for the last one, run one task on each 
available core in the platform. If $m$ tasks are scheduled on the same compute 
node then they read input, resp. write output, with a data rate $bw_{read}/m$, 
resp. $bw_{write}/m$. For all but the last batch of tasks, $m=p$, while for the 
last batch of tasks $m \leq p$. In this way, we obtain an execution time estimate 
for each task in a batch. Since batch executions can overlap (i.e., after the 
shortest task in a batch completes on a core, a task in the next batch starts 
executing on that core), we estimate the execution time of a batch as the average 
execution time of its tasks. 

Writing a closed-form formula for this model is tedious, but it is 
straightforward to implement it programmatically~\cite{wfbench-github}.  
Note that this model is 
likely more sophisticated than what a typically user may develop.

\subsection{Model Evaluation}

\begin{figure}[!t]
    \centering
    \includegraphics[width=\linewidth]{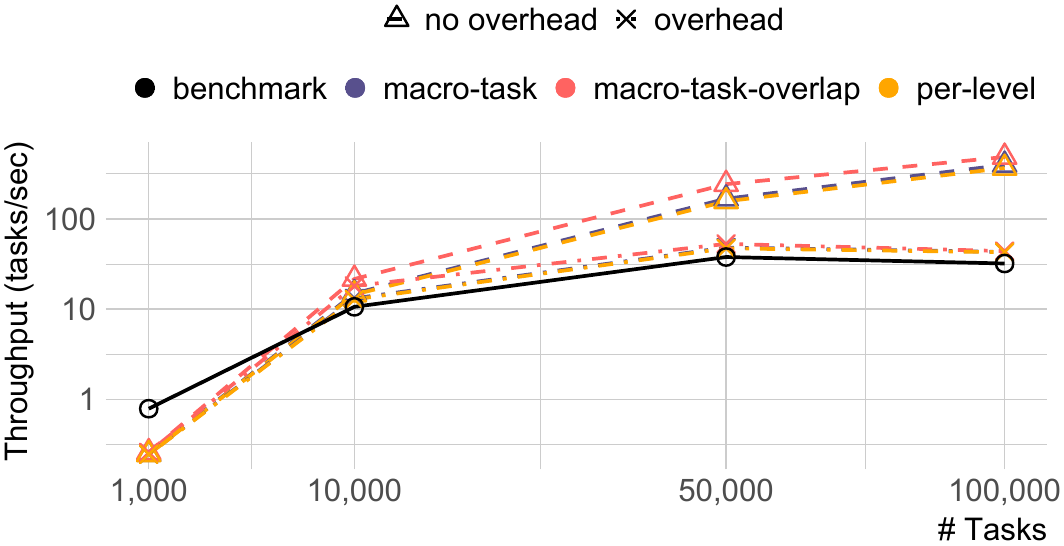}
    \vspace{-20pt}
    \caption{Throughput vs. number of tasks for the Blast application with 500GB data footprint.}
    \vspace{-10pt}
    \label{fig:model:blast-500}
\end{figure}
\begin{figure}[!t]
    \centering
    \includegraphics[width=\linewidth]{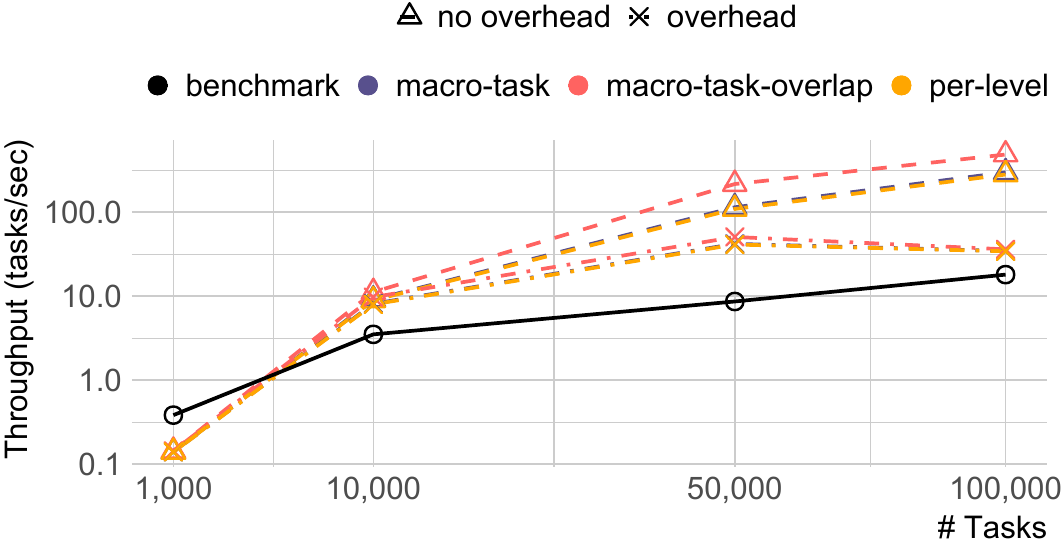}
    \vspace{-20pt}
    \caption{Throughput vs. number of tasks for the Montage application with 1000GB data footprint.}
    \vspace{-10pt}
    \label{fig:model:montage-1000}
\end{figure}

Figures~\ref{fig:model:blast-500} and~\ref{fig:model:montage-1000} show 
throughput vs. number of tasks as measured based on benchmark executions and 
as estimated using the three above models, for 
Blast and Montage with 500~GB and 1~TB data footprints, 
respectively. Results are similar for other applications and/or data footprint 
combinations.  The relative throughput of the three estimates is as expected, 
with the macro-task estimate that assumes full overlap of I/O and computation 
leading to higher throughput than its no-overlap counterpart, and the per-level 
estimate leading to the lowest throughput, as it accounts for the 
workflow's structure. The main observation, however, is that there is a 
large discrepancy between the estimated and the benchmarked throughput (up 
to one order of magnitude for 100,000 tasks).  

\begin{figure}[!t]
    \centering
    \includegraphics[width=\linewidth]{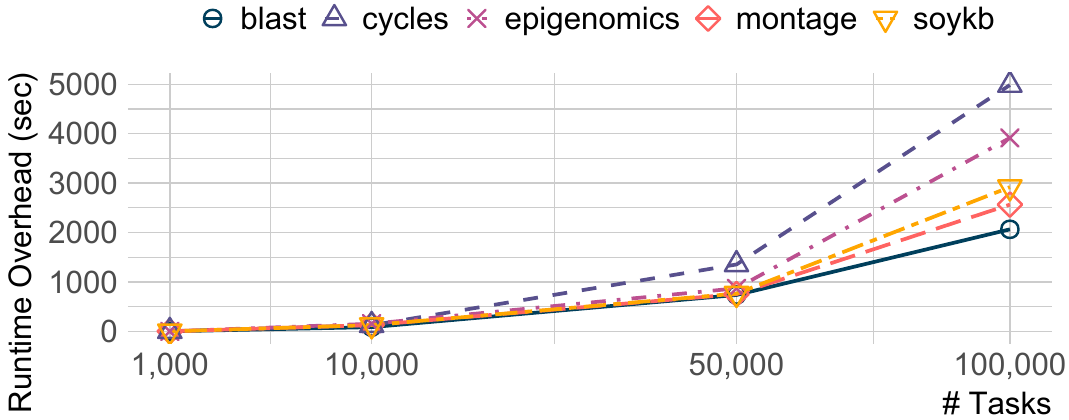}
    \vspace{-20pt}
    \caption{Workflow execution time (or total overhead) vs. number of tasks.}
    \vspace{-10pt}
    \label{fig:summit-overhead}
\end{figure}

The main reason for the observed discrepancy is that our models do not account 
for the overhead of the workflow system. To measure this overhead, we 
executed our benchmarks on Summit with zero data footprint and where all workflow 
tasks perform zero work. Figure~\ref{fig:summit-overhead} shows measured 
makespans for these executions.  The general and expected trend is that overhead 
increases as the number of tasks increases.  But overhead behavior differs across 
applications and developing an accurate overhead model may not be 
straightforward. Instead, we simply add measured total overhead for each 
application and number of tasks to the makespan estimates computed using the 
models in the previous section.  Throughput values computed using these modified 
estimated makespans are shown in Figures~\ref{fig:model:blast-500} 
and~\ref{fig:model:montage-1000} as dotted lines, and are much closer to the 
measured throughput for the benchmark executions. The per-level model, augmented with 
measured overhead, leads to the most accurate throughput estimates.

\begin{figure}[!t]
    \centering
    \includegraphics[width=\linewidth]{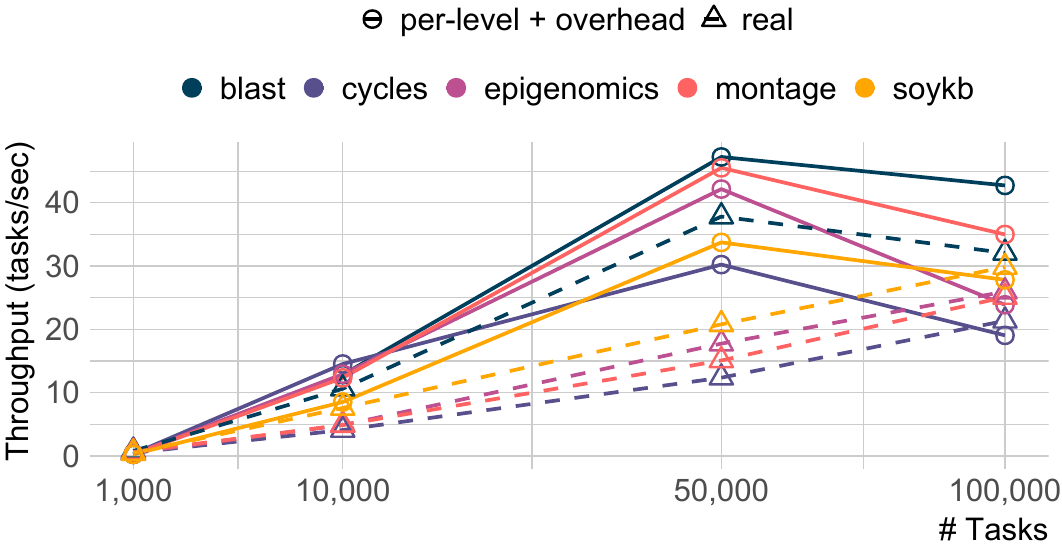}
    \vspace{-20pt}
    \caption{Throughput vs. number of tasks for all applications with 500~GB 
            footprint. Solid, resp. dashed, lines for measured, vs. estimated, 
            makespans. Estimates are computed using the per-level model with 
            added measured overhead.}
    \vspace{-10pt}
    \label{fig:overall-estimate-500-overhead}
\end{figure}

One may wonder whether throughput estimates computed using the per-level model 
(with added measured overhead) paint an accurate picture of the relative 
performance of the different workflow applications.  
Figure~\ref{fig:overall-estimate-500-overhead} shows measured and estimated 
throughput vs. number of tasks for all workflow applications with a 500~GB data 
footprint.  There are clear differences between the measured and estimated 
throughput, both in terms of the relative ranking of the applications and in the 
trends. For instance, while the measured throughput shows that the Montage 
application achieves the second lowest throughput across all five applications, 
the estimated throughput instead shows it to achieve the second highest 
throughput. For 50,000 tasks the estimates show Blast and Montage to achieve 
similar throughput, while measurements show more than a factor 2 difference. 
Furthermore, while in the measured results all applications but Blast show 
monotonically increasing throughput as the number of tasks increases, the 
throughput estimates show non-monotonically increasing throughput for all 
applications.  Results are similar for the 100~GB and 1~TB data footprints, in 
that there are clear discrepancies between measured and estimated relative 
throughput and their trends. These discrepancies are slightly larger when using 
the macro-task models, and much larger when not adding the measured overhead to 
the estimated makespans.

We conclude that estimating workflow performance by composing I/O and compute 
benchmark results on a particular platform using the above models 
does not paint an accurate picture of workflow performance, even assuming 
that the overhead behavior of the workflow system is perfectly known 
(which is not true in practice). While developing accurate models may be 
possible, doing so, especially so that they remain accurate across platform 
configurations, workflow applications and workflow systems, is likely 
a very steep challenge. This justifies the need for and the usefulness of the workflow 
benchmarks developed in this work.

\section{Conclusion}
\label{sec:conclusion}

We have presented an approach for automatically generating realistic workflow 
benchmark specifications that are easily translated into benchmark code that 
is executable with current workflow systems. This approach generates 
workflow tasks with arbitrary performance characteristics regarding CPU, 
memory, and I/O usage, and generates realistic task dependency structures 
based on that seen in real-world scientific workflow applications.  We have 
presented experimental results that show that our approach can be used to 
instantiate benchmarks whose executions and execution performance are 
representative of production workflow applications. We have also conducted a 
case-study that demonstrates the use and the usefulness of our generated 
benchmarks.

One future work direction is to extend our generated benchmarks to support 
other or emerging workflow scenarios, e.g., scenarios in which workflow tasks 
are MPI-based parallel programs (e.g., ``HPC 
workflows"~\cite{mattoso2013user}). We also plan to support GPU workflow 
task benchmark implementation so that users can specify memory work on the GPU 
(for global and shared memory), define data transfer from/to the GPU, or use 
an already available GPU benchmark. 
Finally, we wish to extend our approach to generate benchmarks that perform 
``in situ" executions, that is, where intermediate workflow data can remain 
in memory rather than being stored on disk.

{
\small
\medskip
\noindent \textbf{Acknowledgments.}
This work is funded by NSF contracts \#2106059, \#2106147, \#2103489,  
\#2103508, \#1923539, and \#1923621, and supported by ECP (17-SC-20-SC), 
a collaborative effort of the U.S. 
DOE Office of Science and the NNSA.
This research used resources of the OLCF at ORNL, which is supported by 
the Office of Science of the U.S. DOE under Contract No. DE-AC05-00OR22725.
We thank the NSF Chameleon Cloud for providing access to their 
resources. 
}


\balance

\end{document}